\documentclass[prd, aps, twocolumn, reprint, amsfonts, amssymb, amsmath, preprintnumbers, showpacs, nofootinbib, superscriptaddress]{revtex4}
\usepackage{graphicx, multirow,soul}
\usepackage[citecolor=blue]{hyperref}
\usepackage[all]{hypcap}
\usepackage{url}
\usepackage{color}
\usepackage{bbold}
\usepackage{ulem}
\usepackage{orcidlink}
\usepackage{array}
\usepackage[compact]{titlesec}
\titlespacing{\section}{0pt}{2ex}{1ex}

\usepackage{comment}
\usepackage{rotating}      
\usepackage{multirow}      
\usepackage{booktabs}     
\usepackage{graphicx}      
\usepackage{amsmath}       
\usepackage{siunitx}       

\allowdisplaybreaks

\begin{document}

\title{Probing quark-lepton correlation in GUTs with high-precision neutrino measurements}

\author{Zi-Qiang Chen \orcidlink{0009-0004-9856-6682}}
%\email[Email:\;]{chenziqiang22@mails.ucas.ac.cn}
\affiliation{School of Fundamental Physics and Mathematical Sciences,\\
			\it Hangzhou Institute for Advanced Study, UCAS, Hangzhou 310024, China}
\affiliation{Institute of Theoretical Physics, Chinese Academy of Sciences, Beijing 100190, China}
\affiliation{University of Chinese Academy of Sciences, Beijing 100049, China}
\author{Gao-Xiang Fang \orcidlink{0009-0008-5785-6690}}
%\email[Email:\;]{fanggaoxiang21@mails.ucas.ac.cn}
\affiliation{School of Fundamental Physics and Mathematical Sciences,\\
			\it Hangzhou Institute for Advanced Study, UCAS, Hangzhou 310024, China}
\affiliation{Institute of Theoretical Physics, Chinese Academy of Sciences, Beijing 100190, China}
\affiliation{University of Chinese Academy of Sciences, Beijing 100049, China}
\author{Ye-Ling Zhou \orcidlink{0000-0002-3664-9472}}
%\email[Email:\;]{zhouyeling@ucas.ac.cn}
\affiliation{School of Fundamental Physics and Mathematical Sciences,\\
			\it Hangzhou Institute for Advanced Study, UCAS, Hangzhou 310024, China}

\date{\today}

\begin{abstract}
GUTs unify quarks and leptons into same representations and predict correlations between their masses and mixing. 
We perform numerical scans in SO(10) GUTs to explore the flavor space with new data of JUNO taken into account. 
The quark-lepton correlation shows the preference of normal ordering for light neutrino masses, predicts favored region of the CP-violating phase in neutrino oscillations, and classifies GUT models based on their testability in neutrinoless double beta decay experiments.
The quark-lepton correlation predicts mass spectrum of right-handed neutrinos, pointing to the energy scale of baryon and lepton number violation and providing sources for baryogenesis.
We emphasize that, as high precision measurements of neutrino physics are  coming, the quark-lepton correlation will provide increasingly important role in the testability of GUTs, complementary to proton decay measurements. 
\end{abstract}

%\preprint{} 
% \pacs{}
\maketitle
%%%%%%%%%%%%%%%%%%%%%%%%%%%%%%%%%%%%%%%%%%%%%%%%%%%%%%%%%%%
%%%%%%%%%%%%%%%%%%%%%%%%%%                          Introduction                             %%%%%%
%%%%%%%%%%%%%%%%%%%%%%%%%%%%%%%%%%%%%%%%%%%%%%%%%%%%%%%%%%%

\section{INTRODUCTION}

Grand Unified Theories (GUTs) do not just unify three gauge interactions of the Standard Model (SM), but also unify Yukawa interactions of quarks and leptons. 
Of particular interest are the $SO(10)$ GUTs~\cite{Fritzsch:1974nn} where all SM matter fields are arranged in $16$-plets of $SO(10)$ and, as a consequence, all fermion masses and mixing are correlated \cite{Chanowitz:1977ye,Ramond:1979py,Harvey:1981hk}. 
While the proton decay provides a smoking gun signature of GUTs \cite{Weinberg:1979sa,Wilczek:1979hc,Weinberg:1980bf,Weinberg:1981wj,Sakai:1981pk,Dimopoulos:1981dw,Ellis:1981tv}, the quark-lepton correlation of masses and mixing might verify the unification of matters via the Yukawa sector.
Furthermore, as right-handed neutrinos (RHNs) are embedded into $16$-plets, their mass spectrum and Yukawa couplings are predicted by the quark-lepton correlation, and successful leptogenesis becomes testable in the seesaw framework. 

The historical intertwining between neutrino measurements and GUTs dates back to the early 1980's, when the first generation experiments represented by IMB and KamiokaNDE were operated (see in review \cite{Langacker:1980js}). Current best limits on proton decay are set by the second generation singularly  comprised of Super-Kamiokande \cite{Abe:2014mwa, Miura:2016krn}. The latter is most famous for its observation of atmospheric neutrino oscillation in 1998 \cite{Super-Kamiokande:1998kpq}.

JUNO \cite{An:2015jdp}, Hyper-K \cite{Acciarri:2015uup} and DUNE \cite{Abe:2018uyc} are three main large-scale neutrino oscillation measurements in this stage. They will continue on proton decay measurements to test GUTs and push sensitivities by one order of magnitude. In the meantime, the most important goals are the determination of the neutrino mass ordering (normal $m_1 < m_2 < m_3$ or inverted $m_3 < m_1 < m_2$) and measurement of the Dirac-type CP-violating phase $\delta$. The octant of $\theta_{23}$ (1st octant $0<\theta_{23}<45^\circ$) and (2nd octant $45^\circ<\theta_{23}<90^\circ$) will be confirmed and measurements of other oscillation parameters with precision up to sub-percent level will be carried out.

JUNO released the first data on the measurement of $\sin^2\theta_{12}$ and $\Delta m^2_{21}$ after two months of running, with precision improved by a factor of 1.6 relative to the combination of all previous measurements \cite{JUNO:2025gmd}.
In the stage of high-precision neutrino data, the quark-lepton correlation provides a tool of increasing significance to test GUTs complementary to proton decay. We will use the quark-lepton correlation in $SO(10)$  GUTs as a criterion to explore the preferred distribution of unresolved physical observables, including neutrino mass ordering, the favored range of the Dirac CP-violating phase, and effective mass in neutrinoless double beta ($0\nu\beta\beta$) decay. These quantities will all be tested in the precision era of neutrino physics. In addition, the quark-lepton correlation predicts right-handed neutrino mass spectrum, which touches to the $B-L$ breaking scale and provides sources for baryon-antibaryon asymmetry in the observed Universe. 
The rest of this paper is organized as follows.
Section~\ref{sec:2} gives an overview on Yukawa couplings in $SO(10)$  GUTs.
Numerical analysis based on quark-lepton correlation is performed
in Section~\ref{sec:3}.
We summarize and discuss our results in Section~\ref{sec:4}.

%%%%%%%%%%%%%%%%%%%%%%%%%%%%%%%%%%%%%%%%%%%%%%%%%%%%%%%%%%%
%%%%%%                                        Overview											 %%%%%%
%%%%%%%%%%%%%%%%%%%%%%%%%%%%%%%%%%%%%%%%%%%%%%%%%%%%%%%%%%%
\section{OVERVIEW ON YUKAWAS}  \label{sec:2}

In $SO(10)$  GUTs, all matter fields, including SM fermions (quarks, charged leptons and the left-handed neutrinos) and right-handed neutrinos, of the same generation are arranged into a single 16-dim chiral representation ${ 16}_F$. 
Following the decomposition ${ 16} \times { 16} = { 10} + { 120} + { 126}$, a ``full'' Higgs content for Yukawa couplings could be ${ 10}_H$, ${ 120}_H$ and $\overline{ 126}_H$.  %
Yukawa interactions in $SO(10)$  GUTs are generically given by
\begin{align} \label{eq:general_Yukawas}
{ 16}_F \big[ Y_{10} { 10}_H + Y_{126} \overline{ 126}{}_H + Y_{120} { 120}_H \big] { 16}_F + {\rm h.c.} \,.
\end{align}
where $Y_{10}$, $Y_{126}$ and $Y_{120}$ are $3\times 3$ matrices in the flavor space with absolute values for each entries $\lesssim {\cal O}(1)$ to ensure the perturbativity.
Here, ${ 10}_H$, ${ 120}_H$ can be either real or complex, which will be denoted with superscripts $\mathbb{R}$ and $\mathbb{C}$, respectively, and 
 $\overline{ 126}_H$ is always complex following its representation property. Their roles in fitting fermion data are summarised in Table~\ref{tab:1}.
Complexifying the Higgs induces couplings with the conjugates $({ 10}_H^\mathbb{C})^*$ and $({ 120}_H^\mathbb{C})^*$. They can be forbidden by imposing the global Peccei-Quinn (PQ) symmetry $U(1)_{\rm PQ}$ \cite{Babu:1992ia, Bajc:2005zf}, which gives a compelling explanation of the strong CP problem \cite{Peccei:1977hh}. 
Fermions and Higgses under $U(1)_{\rm PQ}$ transform as
\begin{align}
16_F \to e^{i \alpha} 16_F\,, \hspace{11mm} &
10_H^{\mathbb{C}} \to e^{-2 i \alpha} 10_H^{\mathbb{C}}\,, \nonumber\\
 \overline{126}{}_H^{\mathbb{C}} \to e^{-2 i \alpha} \overline{126}{}_H^{\mathbb{C}}\,,\quad &
120_H^{\mathbb{C}} \to e^{-2 i \alpha} 120_H^{\mathbb{C}}\,. 
\end{align}
Accounting for 3 flavors of fermions, all Yukawa couplings are $3 \times 3$ matrices, and in particular,
$Y_{10}$, $Y_{126}$ are symmetric and $Y_{120}$ is anti-symmetric in the flavor basis.

Following the symmetry breaking from GUTs to the SM, these Higgs fields are eventually decomposed to eight electroweak doublets in total, 
\begin{align} \label{eq:higgses}
{h_i} = \{\tilde{h}_{ 10}^u, h_{ 10}^d, \tilde{h}_{\overline{ 126}}^u, h_{\overline{ 126}}^d, \tilde{h}_{ 120_1}^u, h_{ 120_1}^d, \tilde{h}_{ 120_{15}}^{u}, h_{ 120_{15}}^{d} \} \,,
\end{align}
where $\tilde{h}_{ 10}^u = i \sigma_2 (h_{ 10}^{u})^*$ \cite{Fu:2022lrn}. These field decompositions introduce particles beyond the SM spectrum and the SM Higgs is regarded as the lightest and massless one of their eigenstates. 
Below the GUT scale, it is convenient to write out Yukawa coupling matrices for quarks and leptons 
\begin{align} \label{eq:general_Yukawas_2}
		Y_{u}&= c_{ 10}^u \, Y_{ 10} + c_{ 126}^u Y_{ 126}^{} 
		+ (c_{{ 120}_1}^{u}+c_{{ 120}_{15}}^{u}) \, Y_{ 120} \,,\nonumber\\
		Y_{d}&= c_{ 10}^d \, Y_{ 10}^{} + c_{ 126}^d Y_{ 126}^{} 
		+ (c_{{ 120}_1}^{d}+c_{{ 120}_{15}}^{d}) \, Y_{ 120} \,,\nonumber\\
		Y_{e}&= c_{ 10}^d \, Y_{ 10}^{} - 3 c_{ 126}^d Y_{ 126}^{} 
		+(c_{{ 120}_1}^{d}-3c_{{ 120}_{15}}^{d}) \,Y_{ 120} \,,\nonumber\\
		Y_{\nu}&= c_{ 10}^u \, Y_{ 10}^{} - 3 c_{ 126}^u Y_{ 126}^{} 
		+ (c_{{ 120}_1}^{u}-3c_{{ 120}_{15}}^{u}) \, Y_{ 120} \,.
\end{align}
And mass matrices for quarks and leptons as well as the Dirac mass matrix for neutrinos are generated after the electroweak symmetry breaking, $M_f = Y_f v$ with $f = u,d,e,\nu$ and $v = 174$~GeV. 
The $\overline{126}_H$ Higgs contains scalars of the charge $B-L=2$. They are essential to generate light neutrino masses via seesaw mechanism, 
\begin{align}
	M_R &= v_S Y_{126} \,, \nonumber\\
	M_\nu &= - Y_\nu M_R^{-1} Y_\nu^T v^2 \,,
\end{align}
where $v_S$ is the scalar VEV and we have assumed type-I  seesaw dominance. %
Since all SM fermion mass matrices $M_u, M_d, M_e$ and $M_\nu$ are correlated, it is obvious that masses of quarks and leptons, and the consequent flavor mixing (i.e., Cabibbo-Kobayashi-Maskawa (CKM) and Pontecorvo-Maki-Nakagawa-Sakata (PMNS) mixing), are all correlated via the above formulas. 

 \begin{table}[t!] 
	\renewcommand{\arraystretch}{1.3}
	\centering
\caption{A complete list of possible combinations of Higgs content in renormalizable $SO(10)$  GUTs with all fermions arranged in ${16}_F$.}
	\label{tab:1}	
	\begin{tabular}{lll}		
	\hline
		Higgs content & Whether realistic in fitting fermion data & \\
		\hline
		Only one Higgs & No, no flavor mixing\\

		${ 10}_{H}^\mathbb{R,C}, { 120}_{H}^\mathbb{R,C}$ & No,  massless light neutrinos\\
		
		${ 10}_{H}^\mathbb{R}, \overline{ 126}_{H}^\mathbb{C}$ & No, small hierarchy between $m_b$ and $m_t$ \\

		${ 120}_{H}^\mathbb{R,C}, \overline{ 126}_{H}^\mathbb{C}$ & No, inconsistent with mixing data \\
		
		${ 10}_{H}^\mathbb{C}, \overline{ 126}_{H}^\mathbb{C}$ & Yes, imposing $U(1)_{\rm PQ}$ leads to {\tt M1} \\

		${ 10}_{H}^\mathbb{R}, \overline{ 126}_{H}^\mathbb{C}, { 120}_{H}^\mathbb{R}$ & Yes, assuming SCPV leads to {\tt M2} \\
		${ 10}_{H}^\mathbb{C}, \overline{ 126}_{H}^\mathbb{C}, { 120}_{H}^\mathbb{C}$ & Yes, imposing $U(1)_{\rm PQ}\times Z_2^P$ leads to  {\tt M3} \\
		\hline
	\end{tabular}
\end{table}

While the general Yukawa couplings in Eq.~\eqref{eq:general_Yukawas} leave plenty of room to fit data of fermion masses and mixing, economical models have been considered to reduce the number of free parameters. We discuss the economical and realistic Higgs content to be included in this work. 
First, one single Higgs multiplet (either ${ 10}_H$, $\overline{ 126}_H$ or ${ 120}_H$) is not enough, since it fails to generate  fermion masses and no flavor mixing in either quark or lepton sector.
To explain tiny masses for light neutrinos, $\overline{126}_H$ must be included since it is the only Higgs involving a SM singlet with charge $B-L=2$, which is necessary to generate Majorana masses for neutrinos via seesaw mechanism.
The minimal extension is by adding a real $10^{\mathbb{R}}_H$ into the theory. However, it cannot explain the large hierarchy between $m_b$ and $m_t$ \cite{Bajc:2005zf}. On the other hand, the model with $(\overline{126}_H, 120_H)$ is found to be unable to fit all the fermion masses and mixing angles in the three generation case \cite{Joshipura:2011nn,Dueck:2013gca}. 
Then, we are left with a few realistic choices for fitting data, either the minimal Higgs content $({ 10}_H^{\mathbb{C}}, \overline{ 126}_H^{\mathbb{C}})$ or a complete list $({ 10}_H, \overline{ 126}_H, { 120}_H)$. Among them, we will focus on three compelling models, which are summarised in Table~\ref{tab:2}. \\[2mm]

%%%%%%%%%%%%%%%%%%%%%%%%%%%%%%%%%%%%%%%%%%%%%%%%%%%%%%%%%%%
%%%%%%                               Models												 %%%%%%
%%%%%%%%%%%%%%%%%%%%%%%%%%%%%%%%%%%%%%%%%%%%%%%%%%%%%%%%%%%
%%%%%%%%%%%%%%%%%%%%%%%%%%%%%%%%%%%%%%%%%%%%%%%%%%%%%%%%%%%
\begin{table}[t!] 
	\renewcommand{\arraystretch}{1.3}
	\centering
	\caption{Representative models to be used in fitting fermion data, their main features and number of free parameters.}
	\label{tab:2}	
	\begin{tabular}{ccccc}		
	\hline
		Model & Main features & $n_{\rm para}$\\
		\hline
		{\tt M1} & Only $Y_{10}, Y_{126}$  & 19 \\

		{\tt M2} & Real $Y_{10}, Y_{126}, Y_{120}$ & 19\\

		{\tt M3} & Hermitian $Y_u,Y_d,Y_e,Y_\nu$ & 18\\
		\hline
	\end{tabular}
\end{table}

\begin{itemize}

\item {\tt M1} -- {$({10}_H^{\mathbb{C}}, \overline{126}{}_H^{\mathbb{C}})$ with $U(1)_{\rm PQ}$}. \\
\; This one introduces the minimal Higgs contents among the realistic models \cite{Babu:1992ia}. Imposing $U(1)_{\rm PQ}$ forbids terms with $({10}_H^{\mathbb{C}})^*$ and thus only $Y_{10}$ and $Y_{126}$ are present \cite{Mummidi:2021anm,Ohlsson:2019sja,Babu:2015bna,Altarelli:2013aqa}. 
Without loss of generality, we rotate the flavor basis to keep $Y_{126}$ always real and diagonal. There are $6\times 2 + 3$ real parameters in $Y_{10}$ and $Y_{126}$. Additional free parameters includes the relative size and phases among $c^{u,d}_{10,126}$. In total, we have 19 free parameters in the Yukawa sector. 

\item {\tt M2} -- {$({10}_H^{\mathbb{R}}, \overline{126}{}_H^{\mathbb{C}}, {120}_H^{\mathbb{R}})$} with SCPV.\\
Complex Yukawas $Y_{10}$, $Y_{126}$ and $Y_{120}$ leave enough room to fit data. Imposing CP symmetry at high scale reduces the number of free parameters \cite{Babu:2016bmy,Saad:2022mzu,Babu:2024ahk,Babu:2025wop}. In the basis of diagonal $Y_{126}$, there are 6 +3 + 3 real parameters in $Y_{10}$, $Y_{126}$, $Y_{120}$. The Higgses gain complex VEVs and induce spontaneous CP violation (SCPV). 
This model introduce also 19 free parameters in the Yukawa sector competitive to {\tt M1}. 

\item {\tt M3} -- {$({10}_H^{\mathbb{C}}, \overline{126}{}_H^{\mathbb{C}}, {120}_H^{\mathbb{C}})$ with $U(1)_{\rm PQ} \times Z_2^P$}. \\
This model includes a full list of Higgs multiplets \cite{Joshipura:2011nn, Dueck:2013gca, Fu:2022lrn}. $U(1)_{\rm PQ}$ forbids couplings to the complex conjugate of ${10}_H^{\mathbb{C}}$ or ${120}_H^{\mathbb{C}}$. $Z_2^P$ represents the left-right parity symmetry $\psi_L \leftrightarrow \psi_R$, which results in Hermitian Yukawa matrices $Y_u, Y_d, Y_e, Y_\nu$ \cite{Dutta:2004hp}. This symmetry is achieved via a CP parity symmetry at high scale, to guarantee real $Y_{10}$, $Y_{126}$ and $Y_{120}$. Combining the CP parity with the internal matter parity, i.e., the D parity \cite{Chang:1983fu}, $\psi_L \leftrightarrow \psi_R^C$, of $SO(10)$ gives rise to $Z_2^P$. A maximal CP violation is achieved spontaneously if the VEV of ${120}_H^{\mathbb{C}}$ is orthogonal to the other two Higgs VEVs in the complex plane, but $Z_2^P$ is left unbroken \cite{Grimus:2006rk}.

\end{itemize}
We refer to Appendix~\ref{app:1} for detailed parameterizations of these models.

%%%%%%%%%%%%%%%%%%%%%%%%%%%%%%%%%%%%%%%%%%%%%%%%%%%%%%%%%%%
%%%%%%                        Numerical analysis and discussions		             			    			 %%%%%%
%%%%%%%%%%%%%%%%%%%%%%%%%%%%%%%%%%%%%%%%%%%%%%%%%%%%%%%%%%%
\section{NUMERICAL ANALYSIS} \label{sec:3}

We explore the parameter space allowed by current experiments and discuss their physical predictions for the above three models. The  $\chi^2$ function is defined as 
\begin{align}
	\chi^2=\sum_{i=1}^{n_{\rm obs}}\Big(\frac{\mathcal{O}_i-\mathcal{O}^{\rm bf}_i}{\sigma_{\mathcal{O}_i}}\Big)^2 \,,
	\label{eq:chi2}
\end{align} 
where $\mathcal{O}_{i}$ is the $i$-th theoretical prediction, and $\mathcal{O}_i^{\rm bf}$ is the corresponding $i$-th experimental best-fit (bf) value, and the $\sigma_{\mathcal{O}_i}$ is the $1\sigma$ uncertainty.
We take $n_{\rm obs} =18$ observables into fitting, including 9 charged fermion masses, 3 angles and 1 phase in the CKM mixing matrix, 2 mass-squared differences and 3 angles in the PMNS mixing matrix. All data are listed in Appendix~\ref{app:1}. In the neutrino sector, we take NuFIT 6.0 \cite{Esteban:2024eli} but replace data of $\Delta m^2_{21}$ and $\sin^2\theta_{12}$  (best-fits $\pm 1\sigma$) with  the most recent JUNO data,
\begin{align}\label{eq:JUNO}
		&\Delta m^2_{21} = (7.50 \pm 0.12) \times 10^{-5} {\rm eV}^2\,, \nonumber\\
		&\sin^2\theta_{12}^l = 0.3092 \pm 0.0087\,.
\end{align}
NuFIT provides four versions of global data, which we list in Appendix~\ref{app:1} and denote as {\tt N1}, {\tt N2}, {\tt I1} and {\tt I2}, depending on which mass ordering is assumed and whether SK atmospheric data is included. Namely, {\tt N1} and {\tt N2} represent global fits with and without SK atmospheric data in the normal ordering, and {\tt I1} and {\tt I2} are those in the inverted ordering.
The up-to-date experimental constraint on the Dirac CP-violating phase $\delta$ is still weak, and thus is not included in the $\chi^2$. Instead, we treat $\delta$ as a prediction of the model, together with other predictions including the lightest neutrino mass $m_{\text{lightest}}$, the effective masses appearing in $\beta$ decay $m_\beta$, $0\nu\beta\beta$ decay $m_{\beta\beta}$, and right-handed neutrino masses $M_1, M_2, M_3$ (from lightest to heaviest). 

We perform numerical scan for all three models {\tt M1}, {\tt M2}, {\tt M3} with four versions of data respectively accounted. Same procedures of scan are performed in all cases. All viable points for $\chi^2/n_{\rm obs} <10$ in these models are presented in Appendix~\ref{app:1}, in Figs.~\ref{fig:prediction_NO} and \ref{fig:prediction_IO} for normal and inverted ordering, respectively. Representative benchmarks in these models are presented in Appendix~\ref{app:2}.

\begin{figure}[htbp] 
    	\begin{center}     		
	\includegraphics[width=0.48\textwidth]{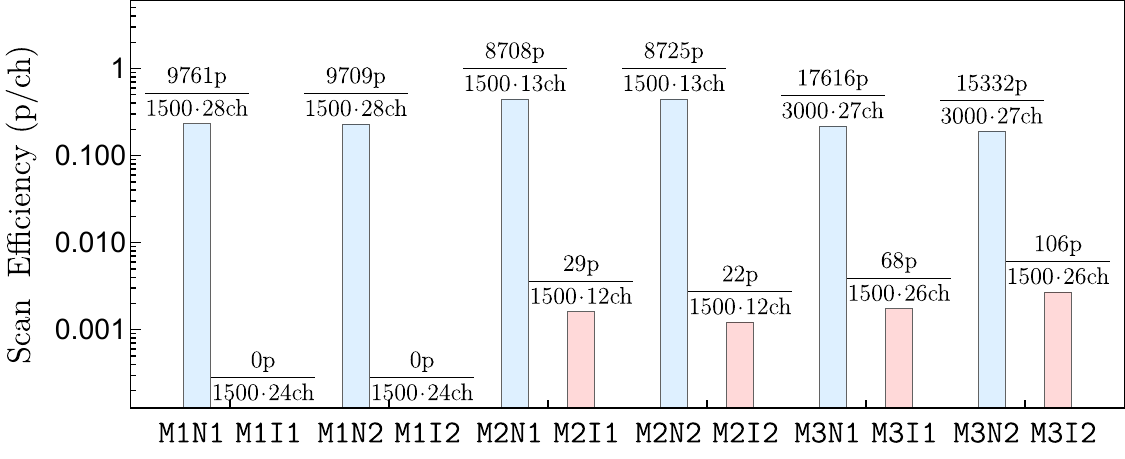}
	\caption{Viable points (p) per core hour (ch) obtained in {\tt M1}, {\tt M2} and {\tt M3} with $\chi^2/n_{\rm obs} < 10$ in each scan.}
    		\label{fig:NOvsIO}
    	\end{center}
    \end{figure}

We first report results on the prediction of neutrino mass ordering. 
In model {\tt M1}, no viable points in the inverted ordering are found to fit data in our scan. In models {\tt M2} and {\tt M3},  although both normal and inverted ordering can fit data, the former case is highly favored due to its efficiency in the scan. Scan efficiencies of models in fitting both mass ordering are compared in Fig.~\ref{fig:NOvsIO}. 

\begin{center}
%%%%%%%%%%%%%%%%%%%%%%%%%%%%%%%%%%%%%%%%%%%%%%%%%%%%%%%%%%%
    \begin{figure}[t!] 
    	\begin{center}     		
	\includegraphics[width=0.9\textwidth]{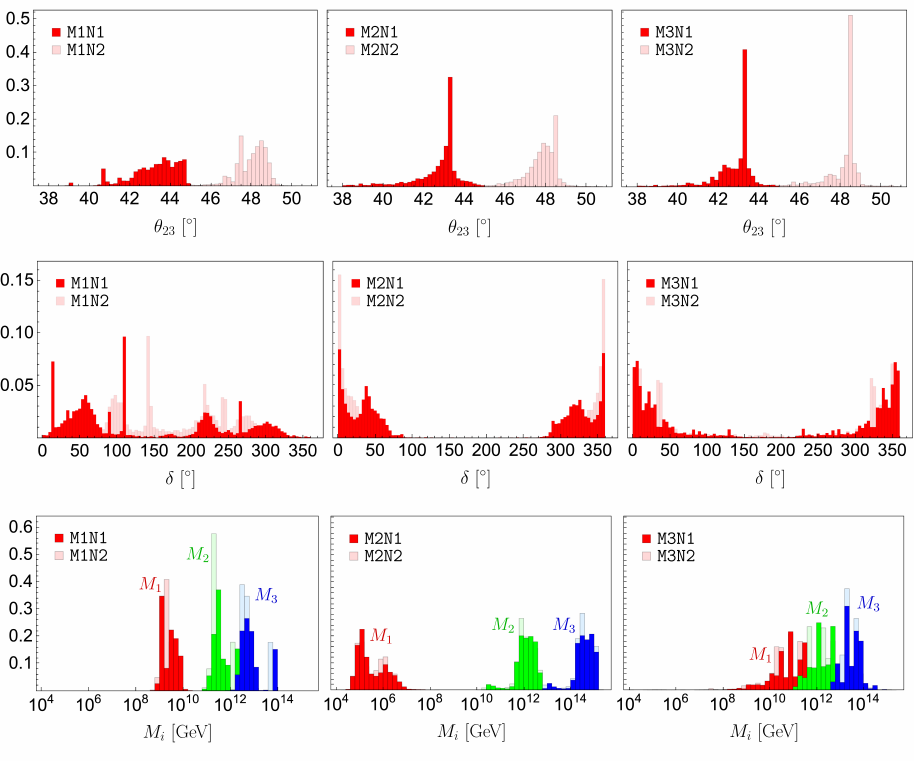}
	\caption{Predicted distributions of some observables of models {\tt M1}, {\tt M2} and {\tt M3} in the normal order in fitting data {\tt N1} and {\tt N2} in the normal ordering. RHN masses $M_1$, $M_2$ and $M_3$ are marked in red, green and blue, respectively in the bottom panel.}
    		\label{fig:NO_weight}
    	\end{center}
    \end{figure}
\end{center}

Below we will mainly focus on the normal mass ordering. Note that data {\tt N1} and {\tt N2} prefer $\theta_{23}$ in the 1st octant ($\theta_{23}<45^\circ$) and 2nd octant ($\theta_{23}>45^\circ$), respectively. Thus, we can further check the preference of the $\theta_{23}$ octant in models. Given points in Fig.~\ref{fig:prediction_NO}, we count points in certain intervals normalized by the corresponding weight $e^{-\chi^2/2}$ and obtain the preferred regimes of observables in models. General results on predictions of $\theta_{23}$, $\delta$ and RHN masses are shown in Fig.~\ref{fig:NO_weight} and summarized as follows.
By taking data of $\theta_{23}$ in both octants, we find a slight preference on the 1st octant in {\tt M2}, but no significant preference of the octant in {\tt M1} and {\tt M3}. 
Distributions of $\delta$ are different in three models. $\delta$ is mainly distributed in the region $[-70^\circ , + 70^\circ]$ in {\tt M2}, extended to $[-90^\circ , + 90^\circ]$ in {\tt M3} and spanned into most of the parameter space in {\tt M1}. 

We discuss the predictions of RHN mass spectrum and potentially consequent phenomenon. These models predict very different mass spectrums for RHNs. We have confirmed the large hierarchy of RHN masses in {\tt M2} which has been studied in \cite{Babu:2024ahk}, where the lightest one can be as small as $10^{4}$~GeV and the heaviest one can reach $10^{15}$~GeV. However, more points show the lightest mass $M_1$ span in the interval $[4\cdot 10^4, 3 \cdot 10^7]$~GeV. This regime has not been discovered before. Both {\tt M1} and {\tt M3}  show much smaller hierarchies. {\tt M1} gives a clear prediction on $M_1$ around $[10^9, 10^{10}]$~GeV and the highest one $M_3\sim [10^{12}, 10^{13}]$~GeV, where most points give $M_1/ M_3 \sim 10^{-3}$. Predictions in {\tt M3} are more diverse, although a lot of them gives $M_1 \sim 10^{10}$~GeV and $M_3\sim 10^{13}$~GeV. This result is consistent with former results \cite{Fu:2022lrn}, but we have found more parameter space.

A clear prediction of RHN mass spectrum serves as the basis for numerous interesting research avenues. We briefly address them below. 
RHNs provide an explanation of baryon-antibaryon asymmetry via the mechanism of leptogenesis. This is quantitatively calculated via the quark-lepton correlation in $SO(10)$  GUTs. A successful thermal leptogenesis requires $M_1$ heavier than $10^9$~GeV \cite{Davidson:2002qv}, which applies for most points in {\tt M1} and {\tt M3}. However, this condition is not satisfied in {\tt M2}, and inspires alternative scenarios, e.g., $N_2$ leptogenesis \cite{DiBari:2010ux}, that might succeed in some parameter space \cite{Babu:2025wop}. 
In a perturbative theory, all RHN masses should not be heavier than the $B-L$ energy scale. That means this scale should be heavier than $M_3 \sim 10^{13}$~GeV for most points in {\tt M1} and {\tt M3}. In {\tt M2}, the heaviest RHN can be as heavy as $10^{15}$~GeV. A $B-L$ scale above this value requires special RG effects to be included in the gauge unification.

\onecolumngrid
%~\newline
%%%%%%%%%%%%%%%%%%%%%%%%%%%%%%%%%%%%%%%%%%%%%%%%%%%%%%%%%%%
\twocolumngrid

%%%%%%%%%%%%%%%%%%%%%%%%%%%%%%%%%%%%%%%%%%%%%%%%%%%%%%%%%%%
%%%%%%                   SUMMARY 					                                						  %%%%%%
%%%%%%%%%%%%%%%%%%%%%%%%%%%%%%%%%%%%%%%%%%%%%%%%%%%%%%%%%%%

\section{CONCLUSION AND DISCUSSIONS}\label{sec:4}

GUTs unify quarks and leptons and predict correlation between their masses and mixing. We emphasize that the quark-lepton correlation can play increasingly important role to test GUTs with the help of high-precision neutrino measurements in the coming decade. 

We begin with a brief review on the fermion masses and mixing in $SO(10)$  GUTs. Then we take three representative GUT models as examples to explore the flavor space consistent with experimental data. In particular, the most recent JUNO data has been taken into account.
The quark-lepton correlation shows the preferred distribution of unresolved physical observables. We argue that inverted ordering for neutrino masses are disfavored in $SO(10)$  GUTs. The Dirac CP-violating phase is predicted differently in these models. Forthcoming measurements will be able to classify GUT models. These quantities will all be tested in the precision era of neutrino physics. 
The quark-lepton correlation further predicts right-handed neutrino masses. Different hierarchies of their mass scales point to different scenarios of leptogenesis in explanation of matter-antimatter asymmetry in our observed universe. The restrictions on the $B-L$ scale  touches to the gauge unification in GUTs.
We suggest to use the quark-lepton correlation as a tool to evaluate GUTs in the precision era of neutrino physics.

%%%%%%%%%%%%%%%%%%%%%%%%%%%%%%%%%%%%%%%%%%%%%%%%%%%%%%%%%%%%%%%%%%%%%%%%%%%%%%%%%%%%%%%%%%%%%%%%%%%%%%%%%%%%%%%%%%%%%%%%%%%%
\acknowledgements
We would like to thank B. Bajc, F. Deppisch and X.H. Hu for useful discussions.
This work was supported by Zhejiang Provincial Natural Science Foundation of China (No.~LDQ24A050002) and 
National Natural Science Foundation of China (NSFC) (No.~12535007 and No.~12547104). 

 \section*{Data Availability}

The code and data that support the findings of this article are openly available (\cite{github:ylz}).
%%%%%%%%%%%%%%%%%%%%%%%%%%%%%%%%%%%%%%%%%%%%%%%%%%%%%%%%%%%%%%%%%%%%%%%%%%%%%%%%%%%%%%%%%%%%%%%%%%%%%%%%%%%%%%%%%%%%%%%%%%%%

\appendix

\section{Details of fitting procedure}\label{app:1}

	After GUT symmetry breaking, ${10}_H,\overline{126}_H,{120}_H$ are decomposed to a series of electroweak doublets of the SM gauge symmetry, denoted in Eq.~\eqref{eq:higgses}. They mix and the lightest massless one matches to the SM Higgs. After the SM Higgs gains the VEV $v$, triggering the electroweak symmetry breaking, the SM fermions obtain masses. We parameterize VEVs of electroweak doublets as
	\begin{align}
		&\{\langle h_{10}^u\rangle,\langle h_{10}^d\rangle,
		\langle h_{120_{1}}^u\rangle, \langle h_{120_{1}}^d\rangle \}\nonumber\\
		=&\{c_{10}^u, c_{10}^d, c_{120_1}^u, c_{120_1}^d \}\times \frac{v}{\sqrt{2}} \,;\nonumber\\
		&\{\langle h_{126}^u\rangle,\langle h_{126}^d\rangle, \langle h_{120_{15}}^u\rangle, \langle h_{120_{15}}^d\rangle\}\nonumber\\
		=&\{c_{126}^u, c_{126}^d, c_{120_{15}}^u, c_{120_{15}}^d\}\times \sqrt{\frac32} v \,,
	\end{align}
	where all $c_{\rm rep}^{u,d}$ are dimensionless and  complex parameters satisfying the normalization 
	\begin{align} 
	&\frac{1}{2} ( |c_{10}^u|^2 + |c_{10}^d|^2 + |c_{120_1}^u|^2 + |c_{120_1}^d|^2) \nonumber\\
	+& \frac{3}{2} ( |c_{126}^u|^2 + |c_{126}^d|^2 + |c_{120_{15}}^u|^2 + |c_{120_{15}}^d|^2 ) = 1,
	\end{align} 
	and overall factors in front of $v$ are induced by representation decomposition in group theory, which are irrelevant to our fit. We further parameterize $c_{\rm rep}^{u,d}$ by removing unphysical parameters. 
	Without loss of generality, $c_{126}^u$ can be rotated to be real, and remaining coefficients are generally complex. Concretely we write
	\begin{align}
		&c_{10}^u=|c_{10}^u|e^{i\alpha},\;c_{10}^d=|c_{10}^d|e^{i\beta},\;\nonumber\\
		&c_{126}^u=|c_{126}^u|,\;c_{126}^d=|c_{126}^d|e^{i\omega},v_S=|v_S|,\nonumber\\
		&c_{120_1}^u+c_{120_{15}}^u=|c_{120_1}^u+c_{120_{15}}^u|e^{i\lambda},\;\nonumber\\
		&c_{120_1}^d+c_{120_{15}}^d=|c_{120_1}^d+c_{120_{15}}^d|e^{i\rho},\nonumber\\
		&c_{120_1}^u-3c_{120_{15}}^u=|c_{120_1}^u-3c_{120_{15}}^u|e^{i\gamma},\nonumber\\
		&c_{120_1}^d-3c_{120_{15}}^d=|c_{120_1}^d-3c_{120_{15}}^d|e^{i\eta},
	\end{align}
Moreover, the SM singlet scalar VEV $v_S$ can also taken to be real.
Therefore, Yukawa matrices and parameters are written as
	\begin{align}
		&D=|c_{126}^u|Y_{126},\;S=|c_{10}^u|Y_{10},\;A=|c_{120_1}^u+c_{120_{15}}^u|Y_{120},\; \nonumber\\
		&v_R=\frac{|v_S|}{|c_{126}^u|},\;r_1=\frac{|c_{126}^d|}{|c_{126}^u|},\;r_2=\frac{|c_{10}^d|}{|c_{10}^u|}, r_3=\frac{|c_{120_1}^d+c_{120_{15}}^d|}{|c_{120_1}^u+c_{120_{15}}^u|},\;
		\nonumber\\
		&r_4=\frac{|c_{120_1}^d-3c_{120_{15}}^d|}{|c_{120_1}^u+c_{120_{15}}^u|},\;r_5=\frac{|c_{120_1}^u-3c_{120_{15}}^u|}{|c_{120_1}^u+c_{120_{15}}^u|}. \label{eq:DSA_vs_Yukawa}
	\end{align}
With the above preparation, three models in this paper can be reparameterized, as below.\\
\begin{itemize}
\item[{\tt M1},] ${10}_H^{\mathbb{C}}+\overline{126}_H^{\mathbb{C}}$ with ${U(1)_{\rm PQ}}$
	\begin{align}
		Y_u&= S+D,\nonumber\\
		Y_d&= r_2 S+e^{i\omega}r_1 D,\nonumber\\
		Y_e&= r_2 S-3e^{i\omega}r_1 D,\nonumber\\
		Y_{\nu}&= S-3D,\nonumber\\
		M_R&=v_R D,
	\end{align}
	where the phase of $c_{\rm 10}^u$, can be absorbed into $S$ and an overall phase between $S$ terms and $D$ terms in $Y_d$ is unphysical and has been ignored. 
	Here we follow the convention in Ref. \cite{Babu:2018tfi}, with a basis where the symmetric matrix $D$ is real and diagonal, whence $S$ becomes a general complex symmetric matrix. $r_1,r_2,v_R$ are real parameters and $\omega$ is a phase.

\item[{\tt M2},] ${10}_H^{\mathbb{R}}+\overline{126}_H^{\mathbb{C}}+{120}_H^{\mathbb{R}}$ with SCPV

	As ${10}_H,{120}_H$ are real fields, VEVs of Higgs doublets in these two fields have relations $c_{10}^{d}=c_{10}^{u*},c_{120_1}^d+c_{120_{15}}^d=c_{120_1}^{u*}+c_{120_{15}}^{u*},c_{120_1}^d-3c_{120_{15}}^d=c_{120_1}^{u*}-3c_{120_{15}}^{u*}$. Next, we follow the convention in Ref. \cite{Babu:2025wop}, the Dirac mass matrices of fermions can be reparameterized as
    \begin{align}
		Y_u&=e^{i\alpha}S+D+e^{i\lambda}A,\nonumber\\
		Y_d&=e^{-i\alpha}S+e^{i\omega}r_1 D+e^{-i\lambda}A,\nonumber\\
		Y_e&=e^{-i\alpha}S-3e^{i\omega}r_1 D+e^{-i\gamma}r_4 A,\nonumber\\
		Y_{\nu}&=e^{i\alpha}S-3D+e^{i\gamma}r_4 A,\nonumber\\
		M_R&=v_R D.
	\end{align}
	Due to the CP symmetry, $Y_{10},Y_{126},Y_{120}$ are all real, equivalently $S,D,A$ are real matrices and $M_R$ is a real symmetric matrix. Here, we work in a basis where $D$ is diagonal and positive. $S$ is real symmetric and $A$ is real antisymmetric. $r_1,r_4,v_R$ are real parameters and $\alpha,\lambda,\gamma,\omega$ are phases.
	
\item[{\tt M3},] ${10}_H^{\mathbb{C}}+\overline{126}_H^{\mathbb{C}}+{120}_H^{\mathbb{C}}$ with ${U(1)_{\rm PQ}}\times{Z_2^P}$
	
	The Dirac mass matrices of fermions are reparameterized as
	\begin{align}
		Y_u&=S+D+iA,\nonumber\\
		Y_d&=r_2S+r_1 D+ir_3A,\nonumber\\
		Y_e&=r_2S-3r_1 D+ir_4 A,\nonumber\\
		Y_{\nu}&=S-3D+ir_5 A,\nonumber\\
		M_R&=v_R D.
	\end{align}
	We follow the convention in Ref. \cite{Fu:2022lrn}, as CP invariance above the GUT scale and SCPV below the GUT scale, $S,D,A$ are real matrices, resulting $Y_u,Y_d,Y_e,Y_\nu$ are Hermitian matrices and $M_R$ is a real symmetric matrix. Consistent with previous models, we still choose a basis where $D$ is diagonal and positive, $S$ is real symmetric and $A$ is real antisymmetric. $r_1,r_2,r_3,r_4,r_5,v_R$ are real parameters. 
\end{itemize}
Parameters of these models are summarized in Table \ref{tab:para}. 

\onecolumngrid
%%%%%%%%%%%%%%%%%%%%%%%%%%%%%%%%%
\begin{center}
	\begin{table}[tbp] 
		\caption{Parameters counting in three models.}
		\label{tab:para}			
		\renewcommand{\arraystretch}{1.3}	
		\centering
		\begin{tabular}{ >{\centering\arraybackslash}m{2cm} >{\centering\arraybackslash}m{11cm} }
			\hline
			Label & Free parameters \\
			\hline
			$\begin{array}{c} {\tt M1} \\ (n_{\rm para}=19) \end{array}$ & 
			$\begin{array}{c}
			\{r_1,r_2\}\in(-10,10), \ m_0\in(0,1)\ {\rm eV}, \ \omega\in[0,2\pi), \\
			D:\{D_{11},D_{22},D_{33}\}\in(0,1), \\
			S:\{S_{11}e^{is_1},S_{12}e^{is_2},S_{13}e^{is_3},S_{22}e^{is_4},S_{23}e^{is_5},S_{33}e^{is_6}\}\in (0,1) e^{i[0,2\pi)}
			\end{array}$ \\
			\hline
			$\begin{array}{c} {\tt M2} \\ (n_{\rm para}=19) \end{array}$ & 
			$\begin{array}{c}
			\{r_1,r_4\}\in(-10,10), ~ m_0\in(0,1)\ {\rm eV}, ~ \{\alpha, \lambda,\gamma,\omega\}\in[0,2\pi), \\ 
			D:\{D_{11},D_{22},D_{33}\}\in(0,1), \\ 
			S:\{S_{11},S_{12},S_{13},S_{22},S_{23},S_{33}\}\in (-1,1), \\ 
			A:\{A_{12},A_{13},A_{23}\}\in (-1,1)
			\end{array}$ \\
			\hline
			$\begin{array}{c} {\tt M3} \\ (n_{\rm para}=18) \end{array}$ & 
			$\begin{array}{c} 
			\{r_1,r_2,r_3,r_4,r_5\}\in(-10,10),~ m_0\in(0,1)\ {\rm eV}, \\ 
			D:\{D_{11},D_{22},D_{33}\}\in(0,1), \\
			S:\{S_{11},S_{12},S_{13},S_{22},S_{23},S_{33}\}\in (-1,1),\\ 
			A:\{A_{12},A_{13},A_{23}\} \in (-1,1)
			\end{array}$ \\
			\hline
		\end{tabular}
	\end{table}
\end{center}
%%%%%%%%%%%%%%%%%%%%%%%%%%%%%%%%%%%%%%%%%%%%%%%%%%%%%%%%%%%

\twocolumngrid

Data ($\text{bf}\pm 1\sigma$) for quark Yukawas and mixing parameters,  as well as charged lepton Yukawas, are listed below, taken from \cite{Babu:2016bmy} with RG effect running up to $M_{\rm GUT}=2\times 10^{16}$ GeV. 
	\begin{align}\label{eq:bfvalues1}
		y_u &= (2.54\pm 0.86) \cdot 10^{-6}, ~
		y_d = (6.56\pm 0.65) \cdot 10^{-6}, \nonumber\\
		y_c &= (1.37\pm 0.04) \cdot  10^{-3}, ~
		y_s = (1.24\pm 0.06)  \cdot 10^{-4}, \nonumber\\
		y_t &=  (0.428\pm 0.003), \hspace{8mm}
		y_b = (5.7\pm 0.005)  \cdot 10^{-3}, \nonumber\\
		\theta^{q}_{12} &= (0.227\pm 0.0006)  \,,\quad\;
		\theta^{q}_{13} = (4.202\pm 0.13)  \cdot  10^{-3}  \,, \nonumber\\
		\theta^{q}_{23} &= (4.858\pm 0.06) \cdot  10^{-2}  \,,\;
		\delta^{q}=(1.207\pm 0.054) \,;
	\end{align}
	\begin{align}
		&
		y_e = (2.703\pm 0.027) \cdot 10^{-6}, ~
		y_\mu = (5.707\pm 0.057) \cdot  10^{-4}, \nonumber\\
		&
		y_\tau = (9.702\pm 0.0097)  \cdot  10^{-3}, 
	\end{align}
For charged lepton Yukawas, we have quoted error to be 1\% uncertainty in their masses to relax the difficulty induced by the tiny  experimental errors of charged lepton masses \cite{Babu:2024ahk}.
Oscillation data except $\Delta m^2_{21}$ and $\sin^2\theta_{12}$ in the normal ordering are taken from NuFIT 6.0 \cite{Esteban:2024eli}. 
We apply following four versions of data: 
normal ordering for IC24 with SK atmospheric data (denoted as {\tt N1}), 
	\begin{align}\label{eq:with_SK}
		\sin^2\theta_{23} &= 0.470 \pm 0.015\,, \;
		\sin^2\theta_{13} = 0.02215 \pm 0.00057 \nonumber\\
		\Delta m^2_{31} &= (2.513 \pm 0.020) \times 10^{-3} {\rm eV}^2 \,;
	\end{align}
normal ordering for IC19 without SK atmospheric data ({\tt N2}),
	\begin{align}\label{eq:without_SK}
		\sin^2\theta_{23} &= 0.561 \pm 0.014\,, \;
		\sin^2\theta_{13} = 0.02195 \pm 0.00056, \nonumber\\
		\Delta m^2_{31} &= (2.534 \pm 0.024) \times 10^{-3} {\rm eV}^2 \,; 
	\end{align}
inverted ordering with SK atmospheric data ({\tt I1}),
	\begin{align}
		\sin^2\theta_{23} &= 0.550 \pm 0.014\,, \;
		\sin^2\theta_{13} = 0.02231 \pm 0.00056\nonumber\\
		\Delta m^2_{32} &= (-2.484 \pm 0.020) \times 10^{-3} {\rm eV}^2 \,;
		\label{eq:with_SK_IO}
	\end{align}
	and 
inverted ordering without SK atmospheric data ({\tt I2}),
         \begin{align}
		\sin^2\theta_{23} &= 0.562 \pm 0.014\,, \;
		\sin^2\theta_{13} = 0.02224 \pm 0.00057 \nonumber\\
		\Delta m^2_{32} &= (-2.510 \pm 0.025) \times 10^{-3} {\rm eV}^2 \,.
		\label{eq:without_SK_IO} 
	\end{align}
In the inverted ordering, both {\tt I1} and {\tt I2} support $\theta_{23}$ in the 2nd octant. 
NuFIT data of $\sin^2\theta_{12}$ and $\Delta m^2_{21}$ in inverted ordering keep the same,
\begin{align}
		\sin^2\theta_{12} &= 0.308 \pm 0.012\,, \;\nonumber\\
		\Delta m^2_{21} &= (7.49 \pm 0.19) \times 10^{-5} {\rm eV}^2 \,, 
\end{align}

In all of our scans, the same procedure is performed as below. 
    \begin{itemize}
	\item[1.] Firstly, we begin with 15000 randomly generated points whose parameters $x_i^0$ 
	(where i runs over all free parameters of each model) are sampled within their defined region 
	shown in Tab. III.  From these parameters, we obtain light neutrino mass matrix via seesaw formula and Dirac mass matrices. The initial value of $\chi^2$ is usually very large, 
	of order $10^8$.
	\item[2.] Then, we minimize $\chi^2$ using the differential evolution (DE) algorithm from the SciPy library. The DE algorithm optimizes all points simultaneously. The configuration uses 15,000 independent chains running in parallel via MPI, with each chain 
	maintaining a population of 15 points. The \texttt{best1bin} mutation strategy ($F=0.7$) is applied for up to 50000 generations.
	\item[3.] Finally, we truncate points with $\chi^2/n_{\rm obs}<10$ for each model.
	
	\end{itemize}

Our scan is performed on the HIAS cluster with 3000 CPU cores available. Code and derived points of all three models are released in GitHub \cite{github:ylz}. We further show these points in Fig.~\ref{fig:prediction_NO} in the normal ordering and Fig.~\ref{fig:prediction_IO} in the inverted ordering. 
We guarantee that all these points are in the perturbative region, e.g., $|Y_{10}|, |Y_{126}|, |Y_{120}| \lesssim {\cal O}(1)$, which is checked via the following way. Once $D$, $S$, $A$ and $r_i$ are obtained in the scan, one can set appropriate values for free parameters $|c_{10}^u|$ and $|c_{126}^u|$ to guarantee the perturbativity of Yukawa couplings $Y_{10} = S / |c_{10}^u|$ and $Y_{126} = D / |c_{126}^u|$. Finally, $Y_{120} = A / |c_{120_1}^u+c_{120_{15}}^u|$ is determined by applying normalization condition of $c^{u,d}_{10,126,120}$ and correlations between $r_i$ and $c^{u,d}_{10,126,120}$ and further one can check the perturbativity of $Y_{120}$. 	

%%%%%%%%%%%%%%%%%%%%%%%%%%%%%%%%%
\onecolumngrid
%%%%%%%%%%%%%%%%%%%%%%%%%%%%%%%%%%%%%%%%%%%%%%%%%%%%%%%%%%%
%%%%%%%%%%%%%%%%%%%%%%%%%%%%%%%%%%%%%%%%%%%%%%%%%%%%%%%%%%%

\begin{center}
%%%%%%%%%%%%%%%%%%%%%%%%%%%%%%%%%%%%%%%%%%%%%%%%%%%%%%%%%%%
    \begin{figure}[htbp] 
    	\begin{center}     		
	\includegraphics[width=0.9\textwidth]{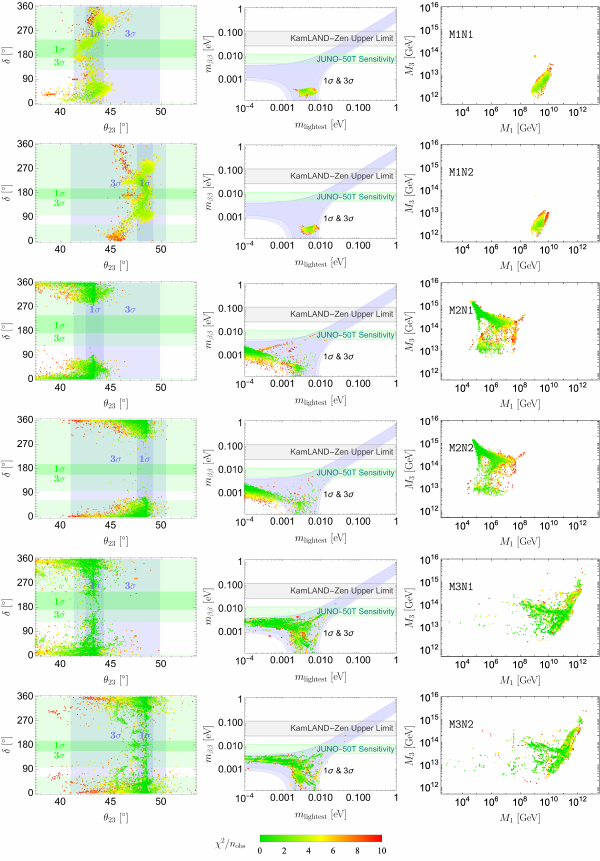}
	\caption{Predictions of models {\tt M1}, {\tt M2} and {\tt M3} for $\chi^2/n_{\rm obs} < 10$ in fitting data {\tt N1} and {\tt N2} in the normal ordering.}
    		\label{fig:prediction_NO}
    	\end{center}
    \end{figure}

%%%%%%%%%%%%%%%%%%%%%%%%%%%%%%%%%%%%%%%%%%%%%%%%%%%%%%%%%%%

%%%%%%%%%%%%%%%%%%%%%%%%%%%%%%%%%%%%%%%%%%%%%%%%%%%%%%%%%%%
\begin{figure}[ttbp] 
	\begin{center}     		
		\includegraphics[width=0.9\textwidth]{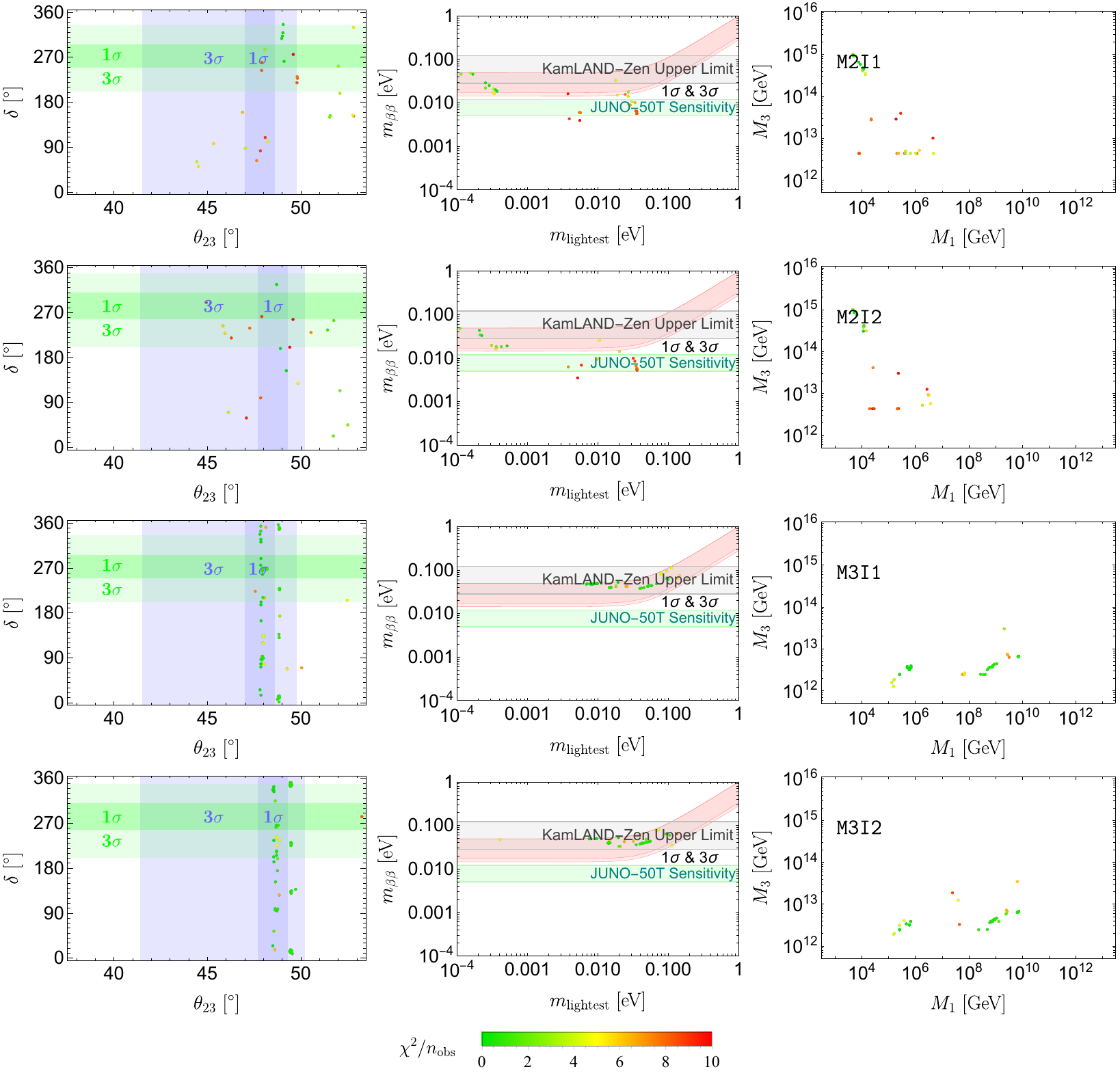}
		\caption{Predictions of models {\tt M2} and {\tt M3} for $\chi^2/n_{\rm obs} < 10$ in fitting data {\tt I1} and {\tt I2} in the inverted ordering. {\tt M1} is not shown since no viable points are found in the inverted ordering.}
		\label{fig:prediction_IO}
	\end{center}
\end{figure}
%%%%%%%%%%%%%%%%%%%%%%%%%%%%%%%%%%%%%%%%%%%%%%%%%%%%%%%%%%%
\end{center}
%%%%%%%%%%%%%%%%%%%%%%%%%%%%%%%%%
\twocolumngrid

In the normal ordering of light neutrino masses, using model {\tt M1} to fit data {\tt N1} ({\tt M1N1}) or {\tt M1} to fit {\tt N2} ({\tt M1N2}) complete their scan using 1500 CPU cores in 28 hours and minimize the $\chi^2$ value to $\chi_{\rm min}^2/n_{\rm obs}=1.54,1.71$. {\tt M2N1} and {\tt M2N2} complete scans with 1500 CPU cores in 13 hours and get $\chi_{\rm min}^2/n_{\rm obs}= 3.34\times 10^{-6}, 7.60\times 10^{-6}$. {\tt M3N1} and {\tt M3N2} complete scans using 3000 CPU cores in 27 hours and get final $\chi_{\rm min}^2/n_{\rm obs} = 3.41\times 10^{-7}, 3.19\times 10^{-7}$. 
	
In the inverted ordering, we fit {\tt I1} and {\tt I2} in model {\tt M1} with 1500 CPU cores in 24 hours, and obtain  $\chi_{\rm min}^2/n_{\mathrm{obs}} = 147.41$ and $147.64$, respectively, thus, no viable point for model {\tt M1} found in the inverted ordering. 
In model {\tt M2}, {\tt M2I1} and {\tt M2I2} complete scans with 1500 CPU cores in 12 hours and obtain $\chi_{\rm min}^2/n_{\mathrm{obs}} = 0.205$ and $0.00721$, respectively.
{\tt M3I1} and {\tt M3I2} complete scans using 1500 CPU cores in 26 hours and obtain $\chi_{\rm min}^2/n_{\mathrm{obs}} = 0.0680$ and $0.598$, respectively.

We comment on the prediction of light neutrino masses which are not discussed in the maintext. In the normal ordering, the lightest neutrino mass $m_1$ is predicted around the region $[0.002, 0.01]$~eV in {\tt M1}, $(0,0.015]$~eV in {\tt M2} and mainly distributed in $(0, 0.02]$~eV in {\tt M3}. $m_{\beta \beta}$ is predicted in a narrow regime less than 0.001~eV  in {\tt M1} and mostly less than 0.003~eV in {\tt M2}, both of which are below the future sensitivity bound. The predicted region in {\tt M3} spans in a wider area, part of which will be tested in JUNO-50T. In the inverted ordering, the lightest neutrino mass $m_3$ is smaller than 0.04~eV in {\tt M2} and smaller than 0.15~eV in {\tt M3}. Some red points predict $m_{\beta\beta}$ less than 0.01~eV due to the prediction of a large $\theta_{12}>40^\circ$. These points should have already been excluded by JUNO since $\theta_{12}$ is far beyond the $3\sigma$ region. However, they are kept in the plot since $\chi^2/n_{\rm obs} <10$ is still satisfied.

%%%%%%%%%%%%%%%%%%%%%%%%%%%%%%%%%
%%%%%%%%%%%%%%%%%%%%%%%%%%%%%%%%%

\section{Benchmark studies} \label{app:2}

\noindent We show benchmark points of {\tt M1}, {\tt M2} and {\tt M3} in Tab.~\ref{tab:benchmark} and \ref{tab:benchmark2} with four versions of data {\tt N1}, {\tt N2}, {\tt I1} and {\tt N2} all taken into account. In models {\tt M2} and {\tt M3}, as diverse mass hierarchies for RHNs are predicted, we show a few distinguishable benchmarks in these models. For example, very large hierarchy $(M_1, M_2, M_3) = (9.95 \cdot 10^{4}, 2.14\cdot 10^{12}, 7.51\cdot 10^{14})$~GeV is predicted in {\tt M2N1B1}, which is consistent with the ansatze assumed in \cite{Babu:2024ahk}, and relatively smaller hierarchy $(M_1, M_2, M_3) = (7.36 \cdot 10^{6}, 6.96\cdot 10^{10}, 2.29\cdot 10^{13})$~GeV is also found in  {\tt M2N1B2} in our scan. We further check the perturbativity of these benchmarks in Table~\ref{tab:Yukawas}. By setting special values of $c_{10}^u$ and $c_{126}^u$, we guarantee that absolute values of all entries of $Y_{10}$, $Y_{126}$ and $Y_{120}$, with the largest entries denoted as ${\rm max}(Y_{10})$, ${\rm max} (Y_{126})$ and ${\rm max} (Y_{120})$, are less than 1. 

\onecolumngrid
%%%%%%%%%%%%%%%%%%%%%%%%%%%%%%%%%
\begin{center}

    \begin{table}[htbp] 
	\small
	\setlength{\tabcolsep}{1.2pt}
	\renewcommand{\arraystretch}{1.2} 
	\caption{Benchmark fit values and predictions in 3 models in the normal ordering. Each benchmark is named, e.g., {\tt M1N1B1}: benchmark 1 of {\tt M1} in fitting data {\tt N1}.  $\Delta m^2_{3l}$ = $\Delta m^2_{31}>0$ and $m_l=m_1$ denote the  lightest neutrino mass. We distinguish different benchmarks, e.g., {\tt M1N1B1} vs {\tt M1N1B2}, due to the different hierarchies of RHN masses  predicted when fitting the same data in the same model.} 
	\label{tab:benchmark}
	\begin{tabular}{c|cc|cccc|cccc}
        \hline
        Physical & \multicolumn{2}{c|}{\textbf{\tt M1}} & \multicolumn{4}{c|}{\textbf{\tt M2}} &     
        \multicolumn{4}{c}{\textbf{\tt M3}} \\
        \cline{2-11}
	Quantity & {\tt M1N1B1} & {\tt M1N2B1} & {\tt M2N1B1} & {\tt M2N1B2} & {\tt M2N2B1} & {\tt M2N2B2} & {\tt M3N1B1} & {\tt M3N1B2} & {\tt M3N2B1} & {\tt M3N2B2} \\
		\hline
		$y_u/10^{-6}$ & 2.65 & 2.75 & 2.55 & 3.07 & 2.53 & 4.85 & 0.637 & 2.54 & 0.744 & 2.54 \\
		$y_c/10^{-3}$ & 1.40 & 1.39 & 1.37 & 1.35 & 1.37 & 1.37 & 1.37 & 1.37 & 1.37 & 1.37 \\
		$y_t$ & 0.429 & 0.427 & 0.428 & 0.429 & 0.428 & 0.430 & 0.428 & 0.428 & 0.428 & 0.428 \\
		$y_d/10^{-6}$ & 3.87 & 3.13 & 6.56 & 7.29 & 6.56 & 7.14 & 5.28 & 6.56 & 5.28 & 6.56 \\
		$y_s/10^{-4}$ & 1.37 & 1.32 & 1.24 & 1.17 & 1.24 & 1.11 & 1.24 & 1.24 & 1.25 & 1.24 \\
		$y_b/10^{-2}$ & 0.571 & 0.571 & 0.570 & 0.570 & 0.570 & 0.568 & 0.569 & 0.570 & 0.569 & 0.570 \\
		$y_e/10^{-6}$ & 2.69604 & 2.70309 & 2.70346 & 2.70190 & 2.70339 & 2.70629 & 2.70318 & 2.70346 & 2.70345 & 2.70341 \\
		$y_{\mu}/10^{-4}$ & 5.65981 & 5.68560 & 5.70708 & 5.72116 & 5.70713 & 5.70725 & 5.71839 & 5.70706 & 5.71432 & 5.70698 \\
		$y_{\tau}/10^{-2}$ & 0.97358 & 0.97054 & 0.97021 & 0.96623 & 0.97020 & 0.97211 & 0.97154 & 0.97020 & 0.97124 & 0.97020 \\
		$\theta_{12}^q$ & 0.22750 & 0.22749 & 0.22739 & 0.22760 & 0.22739 & 0.22721 & 0.22745 & 0.22739 & 0.22745 & 0.22739 \\
		$\theta_{23}^q/10^{-2}$ & 4.904 & 4.872 & 4.858 & 4.876 & 4.858 & 4.835 & 4.858 & 4.858 & 4.856 & 4.858 \\
		$\theta_{13}^q/10^{-3}$ & 4.252 & 4.215 & 4.202 & 4.201 & 4.202 & 4.237 & 4.202 & 4.202 & 4.202 & 4.202 \\
		$\delta^q$ & 1.269 & 1.251 & 1.207 & 1.253 & 1.207 & 1.227 & 1.201 & 1.207 & 1.203 & 1.207 \\
		$\Delta m_{21}^2 / 10^{-5}\,\mathrm{eV}^2$ & 7.53 & 7.50 & 7.50 & 7.50 & 7.50 & 7.48 & 7.50 & 7.50 & 7.50 & 7.50 \\
		$\Delta m_{3\ell}^2 / 10^{-3}\,\mathrm{eV}^2$ & 2.515 & 2.535 & 2.513 & 2.512 & 2.534 & 2.539 & 2.513 & 2.513 & 2.534 & 2.534 \\
		$\mathrm{sin}^2\theta_{12}^l$ & 0.313 & 0.309 & 0.309 & 0.303 & 0.309 & 0.308 & 0.309 & 0.309 & 0.309 & 0.309 \\
		$\mathrm{sin}^2\theta_{23}^l$ & 0.456 & 0.561 & 0.470 & 0.460 & 0.561 & 0.540 & 0.471 & 0.470 & 0.561 & 0.561 \\
		$\mathrm{sin}^2\theta_{13}^l$ & 0.02152 & 0.02180 & 0.02215 & 0.02235 & 0.02195 & 0.02223 & 0.02217 & 0.02215 & 0.02194 & 0.02195 \\
		$\chi^2/(n_{\mathrm{obs}}=18)$ & 1.54 & 1.71 & $5.88\cdot 10^{-6}$ & 0.312 & $7.60\cdot 10^{-6}$ & 0.958 & 0.495 & $3.41\cdot 10^{-7}$ & 0.462 & $3.39\cdot 10^{-7}$ \\
		\hline
		$m_l / \mathrm{eV}$ & $3.34\cdot 10^{-3}$ & $5.60\cdot 10^{-3}$ & $5.29\cdot 10^{-5}$ & $7.81\cdot 10^{-4}$ & $5.41\cdot 10^{-5}$ & $1.52\cdot 10^{-3}$ & $3.82\cdot 10^{-3}$ & $5.46\cdot 10^{-4}$ & $3.44\cdot 10^{-3}$ & $1.03\cdot 10^{-3}$ \\
		$m_{\beta} / \mathrm{eV}$ & $9.40\cdot 10^{-3}$ & $1.05\cdot 10^{-2}$ & $8.85\cdot 10^{-3}$ & $8.88\cdot 10^{-3}$ & $8.85\cdot 10^{-3}$ & $9.02\cdot 10^{-3}$ & $9.64\cdot 10^{-3}$ & $8.87\cdot 10^{-3}$ & $9.49\cdot 10^{-3}$ & $8.91\cdot 10^{-3}$ \\
		$m_{\beta\beta} / \mathrm{eV}$ & $3.67\cdot 10^{-4}$ & $3.85\cdot 10^{-4}$ & $2.47\cdot 10^{-3}$ & $1.40\cdot 10^{-3}$ & $2.54\cdot 10^{-3}$ & $1.17\cdot 10^{-3}$ & $9.06\cdot 10^{-4}$ & $3.15\cdot 10^{-3}$ & $8.36\cdot 10^{-4}$ & $1.70\cdot 10^{-3}$ \\
		$\delta / \mathrm{^\circ}$ & 38.816 & 210.46 & 43.046 & 11.532 & 341.29 & 6.4034 & 186.19 & 14.663 & 189.07 & 351.61 \\
		$M_1 / \mathrm{GeV}$ & $7.48\cdot 10^{9}$ & $2.56\cdot 10^{9}$ & $9.95\cdot 10^{4}$ & $7.36\cdot 10^{6}$ & $8.43\cdot 10^{4}$ & $7.13\cdot 10^{6}$ & $1.01\cdot 10^{3}$ & $1.37\cdot 10^{10}$ & $9.09\cdot 10^{2}$ & $2.71\cdot 10^{10}$ \\
		$M_2 / \mathrm{GeV}$ & $4.95\cdot 10^{11}$ & $2.14\cdot 10^{11}$ & $2.14\cdot 10^{12}$ & $6.96\cdot 10^{10}$ & $3.64\cdot 10^{12}$ & $9.07\cdot 10^{10}$ & $6.36\cdot 10^{11}$ & $2.57\cdot 10^{11}$ & $5.60\cdot 10^{11}$ & $5.01\cdot 10^{11}$ \\
		$M_3 / \mathrm{GeV}$ & $5.61\cdot 10^{12}$ & $2.48\cdot 10^{12}$ & $7.51\cdot 10^{14}$ & $2.29\cdot 10^{13}$ & $1.02\cdot 10^{15}$ & $2.97\cdot 10^{13}$ & $1.36\cdot 10^{13}$ & $9.45\cdot 10^{12}$ & $1.19\cdot 10^{13}$ & $2.09\cdot 10^{13}$ \\
		\hline
	\end{tabular}
\end{table}

%%%%%%%%%%%%%%%%%%%%%%%%%%%
\begin{table}[htbp] 
	\small
	\setlength{\tabcolsep}{1.2pt}
    \renewcommand{\arraystretch}{1.2} 
	\caption{Benchmark fit values and predictions in 2 models in the inverted ordering. Each benchmark is named, e.g., {\tt M2I2B1}: Benchmark 1 of {\tt M2} in fitting data {\tt I2}.  $\Delta m^2_{3 \ell}$ = $\Delta m^2_{32}<0$ and $m_l=m_3$ denote the  lightest neutrino mass.}   	
	\label{tab:benchmark2}
	\begin{tabular}{c|cccc|cccc}
		\hline
		Physical & \multicolumn{4}{c|}{\tt M2 } & \multicolumn{4}{c}{\tt M3 }\\
		\cline{2-9}
		Quantity & {\tt M2I1B1} & {\tt M2I1B2} & {\tt M2I2B1} & {\tt M2I2B2} & {\tt M3I1B1} & {\tt M3I1B2} & {\tt M3I2B1} & {\tt M3I2B2} \\
		\hline
		$y_u/10^{-6}$ & 2.79 & 4.03 & 2.77 & 3.49 & 2.56 & 2.54 & 2.56 & 2.54 \\
		$y_c/10^{-3}$ & 1.37 & 1.44 & 1.37 & 1.36 & 1.38 & 1.36 & 1.38 & 1.36 \\
		$y_t$ & 0.428 & 0.427 & 0.428 & 0.428 & 0.428 & 0.428 & 0.428 & 0.428 \\
		$y_d/10^{-6}$ & 6.92 & 6.82 & 6.60 & 8.92 & 5.95 & 8.05 & 5.95 & 7.93 \\
		$y_s/10^{-4}$ & 1.27 & 0.740 & 1.24 & 1.11 & 1.21 & 1.10 & 1.22 & 1.11 \\
		$y_b/10^{-2}$ & 0.572 & 0.570 & 0.570 & 0.567 & 0.571 & 0.570 & 0.571 & 0.570 \\
		$y_e/10^{-6}$ & 2.70369 & 2.70455 & 2.70332 & 2.70702 & 2.70333 & 2.70326 & 2.70337 & 2.70348 \\
		$y_{\mu}/10^{-4}$ & 5.71365 & 5.73366 & 5.70814 & 5.68800 & 5.70707 & 5.71460 & 5.70701 & 5.71393 \\
		$y_{\tau}/10^{-2}$ & 0.96303 & 0.97624 & 0.96926 & 0.98031 & 0.97013 & 0.97102 & 0.97008 & 0.97048 \\
		$\theta_{12}^q$ & 0.22740 & 0.22733 & 0.22739 & 0.22735 & 0.22742 & 0.22741 & 0.22742 & 0.22740 \\
		$\theta_{23}^q/10^{-2}$ & 4.864 & 4.884 & 4.859 & 4.860 & 4.870 & 4.871 & 4.870 & 4.869 \\
		$\theta_{13}^q/10^{-3}$ & 4.140 & 4.194 & 4.195 & 4.036 & 4.201 & 4.146 & 4.200 & 4.152 \\
		$\delta^q$ & 1.210 & 1.157 & 1.206 & 1.023 & 1.216 & 1.217 & 1.216 & 1.214 \\
		$\Delta m_{21}^2 / 10^{-5}\,\mathrm{eV}^2$ & 7.49 & 7.46 & 7.49 & 7.28 & 7.49 & 7.50 & 7.49 & 7.49 \\
		$\Delta m_{3\ell}^2 / 10^{-3}\,\mathrm{eV}^2$ & $-2.484$ & $-2.483$ & $-2.510$ & $-2.509$ & $-2.484$ & $-2.484$ & $-2.510$ & $-2.510$ \\
		$\mathrm{sin}^2\theta_{12}^l$ & 0.308 & 0.306 & 0.308 & 0.357 & 0.308 & 0.308 & 0.308 & 0.308 \\
		$\mathrm{sin}^2\theta_{23}^l$ & 0.569 & 0.556 & 0.565 & 0.520 & 0.550 & 0.567 & 0.562 & 0.578 \\
		$\mathrm{sin}^2\theta_{13}^l$ & 0.02232 & 0.02228 & 0.02224 & 0.02241 & 0.02231 & 0.02237 & 0.02224 & 0.02230 \\
		$\chi^2/(n_{\mathrm{obs}}=18)$ & 0.205 & 4.31 & 0.00721 & 3.54 & 0.0680 & 0.717 & 0.0680 & 0.598 \\
		\hline
		$m_l / \mathrm{eV}$ & $2.53{\cdot}10^{-4}$ & $1.78{\cdot}10^{-2}$ & $2.07{\cdot}10^{-4}$ & $2.02{\cdot}10^{-2}$ & $9.83{\cdot}10^{-3}$ & $1.48{\cdot}10^{-2}$ & $9.60{\cdot}10^{-3}$ & $1.41{\cdot}10^{-2}$ \\
		$m_{\beta} / \mathrm{eV}$ & $4.88{\cdot}10^{-2}$ & $5.19{\cdot}10^{-2}$ & $4.90{\cdot}10^{-2}$ & $5.31{\cdot}10^{-2}$ & $4.97{\cdot}10^{-2}$ & $5.10{\cdot}10^{-2}$ & $5.00{\cdot}10^{-2}$ & $5.10{\cdot}10^{-2}$ \\
		$m_{\beta\beta} / \mathrm{eV}$ & $2.88{\cdot}10^{-2}$ & $3.30{\cdot}10^{-2}$ & $4.37{\cdot}10^{-2}$ & $1.43{\cdot}10^{-2}$ & $4.92{\cdot}10^{-2}$ & $3.96{\cdot}10^{-2}$ & $4.95{\cdot}10^{-2}$ & $3.95{\cdot}10^{-2}$ \\
		$\delta / \mathrm{^\circ}$ & 307.27 & 101.18 & 326.05 & 69.966 & 327.06 & 347.86 & 202.42 & 14.403 \\
		$M_1 / \mathrm{GeV}$ & $7.60{\cdot}10^{3}$ & $4.73{\cdot}10^{6}$ & $5.12{\cdot}10^{3}$ & $1.84{\cdot}10^{6}$ & $5.90{\cdot}10^{5}$ & $7.09{\cdot}10^{9}$ & $5.93{\cdot}10^{5}$ & $7.14{\cdot}10^{9}$ \\
		$M_2 / \mathrm{GeV}$ & $2.15{\cdot}10^{12}$ & $1.45{\cdot}10^{10}$ & $2.97{\cdot}10^{12}$ & $1.67{\cdot}10^{10}$ & $7.07{\cdot}10^{10}$ & $8.17{\cdot}10^{10}$ & $7.14{\cdot}10^{10}$ & $8.39{\cdot}10^{10}$ \\
		$M_3 / \mathrm{GeV}$ & $6.55{\cdot}10^{14}$ & $4.30{\cdot}10^{12}$ & $8.61{\cdot}10^{14}$ & $5.22{\cdot}10^{12}$ & $3.14{\cdot}10^{12}$ & $6.45{\cdot}10^{12}$ & $3.17{\cdot}10^{12}$ & $6.53{\cdot}10^{12}$ \\
		\hline
	\end{tabular}
\end{table}
\end{center}

    \begin{table}[htbp] 
	\small
	\setlength{\tabcolsep}{1.2pt}
	\renewcommand{\arraystretch}{1.2} 
	\caption{The largest entries of Yukawa coupling matrices $Y_{10}$, $Y_{126}$ and $Y_{120}$ for benchmarks in Tables~\ref{tab:benchmark} and \ref{tab:benchmark2}, where $|c_{10}^u|, |c_{126}^u|$ are set as 0.60, 0.74 for {\tt M1}, and 0.60,0.65 for both {\tt M2} and {\tt M3}, respectively.} 
	\label{tab:Yukawas}
	
	\begin{tabular}{c|cc|cccc|cccc}
		\hline
		$\max(Y)$ & {\tt M1N1B1} & {\tt M1N2B1} & {\tt M2N1B1} & {\tt M2N1B2} & {\tt M2N2B1} & {\tt M2N2B2} & {\tt M3N1B1} & {\tt M3N1B2} & {\tt M3N2B1} & {\tt M3N2B2} \\
		\hline
		$\max(Y_{10})$ & 0.705 & 0.700 & 0.00372 & 0.00951 & 0.00423 & 0.00852  & 0.488 & 0.432 & 0.495 & 0.382 \\
		$\max(Y_{126})$ & 0.0733 & 0.0595 & 0.659 & 0.657 & 0.660 & 0.660 & 0.207 & 0.254 & 0.202 & 0.305 \\
		$\max(Y_{120})$ & --- & --- & 0.0321 & 0.0151 & 0.0337 & 0.0187 & 0.000748 & 0.0623 & 0.000765 & 0.0394 \\
		\hline
		\hline
		 $\max(Y)$ &  &  & {\tt M2I1B1} & {\tt M2I1B2} & {\tt M2I2B1} & {\tt M2I2B2} & {\tt M3I1B1} & {\tt M3I1B2} & {\tt M3I2B1} & {\tt M3I2B2} \\
		 \hline
		 $\max(Y_{10})$ &  &  & 0.00433 & 0.00304 & 0.00439 & 0.00814 & 0.550 & 0.279 & 0.550 & 0.280 \\
		 $\max(Y_{126})$ &  &  & 0.659 & 0.656 & 0.659 & 0.659 & 0.149 & 0.401 & 0.149 & 0.400 \\
		 $\max(Y_{120})$ &  &  & 0.0355 & 0.0157 & 0.0359 & 0.0207 & 0.0312 & 0.0181 & 0.0312 & 0.0179 \\
		 \hline
	\end{tabular}
\end{table}

\twocolumngrid

\end{document}